\documentclass[a4paper]{article}

\usepackage{icrc2013}
\usepackage[english]{babel}
\usepackage{graphicx}
\usepackage{amssymb}
\newcommand{\be}{\,\begin{equation}}
\newcommand{\ee}{\,\end{equation}}
\title{Recent results in cosmic ray physics and their interpretation}

\shorttitle{Cosmic ray physics}

\authors{
Pasquale Blasi$^{1,2}$,
}

\afiliations{
$^1$ INAF/Osservatorio AstroÞsico di Arcetri, Largo E. Fermi, 5, 50125, Firenze, Italy \\
$^2$ INFN/Gran Sasso Science Institute, viale F. Crispi 7, 67100 LÕAquila, Italy
}

\email{blasi@arcetri.astro.it}

\abstract{
The last decade has been dense with new developments in the search for the sources of Galactic cosmic rays. Some of these developments have confirmed the tight connection between cosmic rays and supernovae in our Galaxy, through the detection of gamma rays and the observation of thin non-thermal X-ray rims in supernova remnants. Some other, such as the detection of features in the spectra of some chemicals opened new questions on the propagation of cosmic rays in the Galaxy and on details of the acceleration process. Here I will summarize some of these developments and their implications for our understanding of the origin of cosmic rays. I will also discuss some new avenues that are being pursued in testing the supernova origin of Galactic cosmic rays. 
}


\begin{document}
\maketitle

\section{Introduction}

The 33$^{\rm rd}$ International Cosmic Ray Conference took place one year after the hundredth anniversary of the discovery of Cosmic Rays (CRs). Hence it held sort of a symbolic meaning, in terms of trying to make the point of {\it where we are and we are going} in searching for the sources of CRs. 

\begin{figure}[t]
  \centering
  \includegraphics[width=0.5\textwidth]{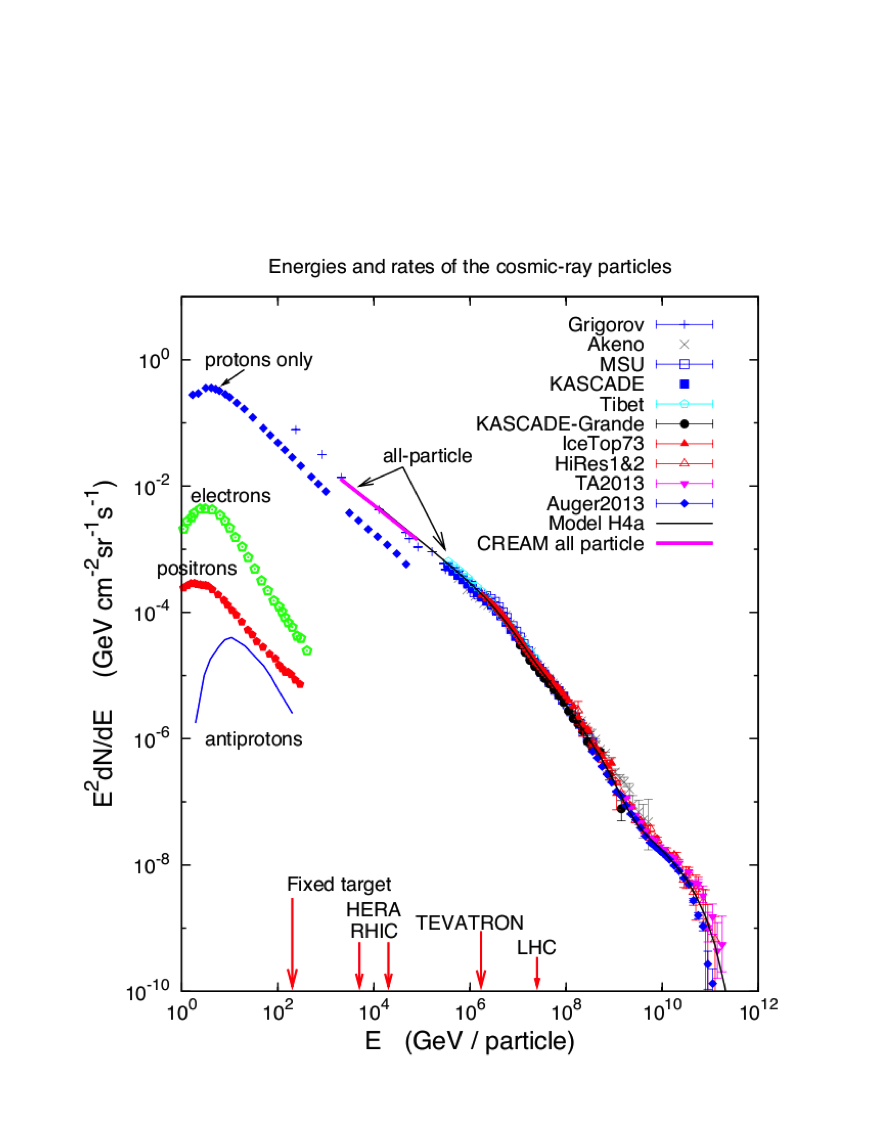}
  \caption{Spectrum of cosmic rays at the Earth (courtesy Tom Gaisser). The all-particle spectrum measured by different experiments is plotted, together with the proton spectrum. The subdominant contributions from electrons, positrons and antiprotons as measured by the PAMELA experiment are shown.}
  \label{fig:spectrum}
 \end{figure}

The flux of all nuclear components present in CRs (the so-called all-particle spectrum) is shown in Fig. \ref{fig:spectrum}. At low energies (below $\sim 30$ GeV) the spectral shape bends down, as a result of the modulation imposed by the presence of a magnetized wind originated from our Sun, which inhibits very low energy particles from reaching the inner solar system. At the knee ($E_{K}=3\times 10^{15}$ eV) the spectral slope of the differential flux (flux of particles reaching the Earth per unit time, surface and solid angle, per unit energy interval) changes from $\sim -2.7$ to $\sim -3.1$. There is evidence that the chemical composition of CRs changes across the knee region with a trend to become increasingly more dominated by heavy nuclei at high energy (see \cite{2006JPhCS4741H} for a review), at least up to $\sim 10^{17}$ eV. At even higher energies the chemical composition remains matter of debate. Recent measurements carried out with KASCADE-GRANDE \cite{Apel:2013p3150} reveal an interesting structure in the spectrum and composition of CRs between $10^{16}$ and $10^{18}$ eV: the collaboration managed to separate the showers in electron-rich (a proxy for light chemical composition) and electron-poor (a proxy for heavy composition) showers and showed that the light component (presumably protons and He, with some contamination from CNO) has an ankle like structure at $10^{17}$ eV. The authors suggest that this feature signals the transition from Galactic to extragalactic CRs (in the light nuclei component). The spectrum of Fe-like CRs continues up to energies of $\sim 10^{18}$ eV, where the flux of Fe and the flux of light nuclei are comparable. Similar results were recently put forward by the ICETOP collaboration \cite{Aartsen:2013p042004}. This finding does not seem in obvious agreement with the results of the Pierre Auger Observatory \cite{2010PhRvL104i1101A}, HiRes \cite{2007JPhG34401S} and Telescope Array \cite{2013EPJWC5206002S}, which find a chemical composition at $10^{18}$ eV that is dominated by the light chemical component. 

The presence of a knee and the change of chemical composition around it have stimulated the idea that the bulk of CRs originates within our Galaxy. The knee could for instance result from the superposition of cutoffs in the different chemicals as due to the fact that most acceleration processes are rigidity dependent: if protons are accelerated in the sources to a maximum energy $E_{p,max}\sim 5\times 10^{15}$ eV, then an iron nucleus will be accelerated to $E_{Fe,max}=26E_{p,max}\sim (1-2)\times 10^{17}$ eV (it is expected that at such high energies even iron nuclei are fully ionized, therefore the unscreened charge is $Z=26$). A knee would naturally arise as the superposition of the cutoffs in the spectra of individual elements (see for instance \cite{Horandel:2004p1543,Blasi:2012p2051,Gaisser:2013p3152}). 

The apparent regularity of the all-particle spectrum in the energy region below the knee is at odds with the recent detection of features in the spectra of individual elements, most notably protons and helium: the PAMELA satellite has provided evidence that both the proton and helium spectra harden at $230$ GV \cite{Adriani:2011p1893}. The spectrum of helium nuclei is also found systematically harder than the proton spectrum, through only by a small amount. The slope of the proton spectrum below $230$ GeV was measured to be $\gamma_{1}=2.89\pm 0.015$, while the slope above $230$ GeV becomes $\gamma_{2}=2.67\pm 0.03$. The slopes of protons and helium spectra at high energies as measured by PAMELA appear to be in agreement with those measured by the CREAM experiment  \cite{Ahn:2010p624} at supra-TeV energies. Some evidence also exists for a similar hardening in the spectra of heavier elements \cite{Ahn:2010p624}. 


During this ICRC, some preliminary data from the AMS-02 experiment on the International Space Station have been presented (see presentation by S. Ting). This data do not confirm the existence of the spectral breaks in the protons and helium spectra, as observed by PAMELA. Given the preliminary nature of these data and the lack of refereed publications at the time of writing of this paper, I cannot comment further on their relevance.

The measurement of the ratio of fluxes of some nuclei that can only be produced by CR spallation and the flux of their parent nuclei provides the best estimate so far of the amount of matter that CRs traverse during their journey through the Galaxy. In order to account for the observed B/C ratio, CRs must travel for times that exceed the ballistic time by several orders of magnitude before escaping the Galaxy (this number decreases with energy). This is the best argument to support the ansatz that CRs travel diffusively in the Galactic magnetic field \cite{1972PhRvL29445J}. A similar conclusion can be drawn from the observed flux of some unstable isotopes such as $^{10}Be$ \cite{Simpson:1988p773}. The decrease of the B/C ratio with energy per nucleon is well described in terms of a diffusion coefficient that increases with energy. 

In principle a similar argument can be applied to the so-called positron fraction, the ratio of fluxes of positrons and electrons plus positrons, $\Phi_{e^{+}}/(\Phi_{e^{+}}+\Phi_{e^{-}})$, where however special care is needed because of the important role of energy losses for leptons. In first approximation, it is expected that positrons may only be secondary products of inelastic CR interactions that lead to the production and decays of charged pions. In this case it can be proved that the positron fraction must decrease with energy. In fact several past observations, and most recently the PAMELA measurements \cite{Adriani:2009p787} and the AMS-02 measurement \cite{Aguilar:2013p3117}, showed that the positron fraction increases with energy above $\sim 10$ GeV. This anomalous behaviour is not reflected in the flux of antiprotons \cite{Adriani:2008p641}: the ratio of the antiprotons to proton fluxes $\Phi_{\bar p}/\Phi_{p}$ is seen to decrease, as expected based on the standard model of diffusion. Although the rise of the positron fraction has also been linked to dark matter annihilation in the Galaxy, there are astrophysical explanations of this phenomenon that can account for the data without extreme assumptions (see the review paper by \cite{2012APh392S} for a careful description of both astrophysical models and dark matter inspired models). 

The simple interpretation of the knee as a superposition of the cutoffs in the spectra of individual elements, as discussed above, would naively lead to the conclusion that the spectrum of Galactic CRs should end at $\sim 26 E_{K}\lesssim 10^{17}$ eV. Clearly this conclusion is not straightforward: some rare type of sources may in principle be able to accelerate CRs to larger energies while leaving the interpretation of the knee unaffected, though changing the energy at which Galactic CRs end. This opens the very important question of where should one expect the transition to extragalactic CRs to take place. Although in the present review I will only occasionally touch upon the problem of ultra high energy cosmic rays (UHECRs), it is important to realize that the quest for their origin is intimately connected with the nature of the transition from Galactic CRs to UHECRs.

At the time of the writing, there is rather convincing and yet circumstantial evidence that the bulk of CRs are accelerated in supernova remnants (SNRs) in our Galaxy, as first proposed by \cite{Baade:1934p886,Ginzburg:1961p170}. The evidence is based on several independent facts: gamma rays unambiguously associated with production of neutral pions have been detected from several SNRs close to molecular clouds \cite{Ackermann:2013p3110,Tavani:2010p2219}; the gamma ray emission detected from the Tycho SNR \cite{2012ApJ744L2G,2011ApJ730L20A} also appears to be most likely of hadronic origin \cite{Morlino:2011p1965,2013ApJ76314B}; the bright X-ray rims detected from virtually all young SNRs (see \cite{Vink:2012p2755,2006AdSpR.37.1902B} for a recent review) prove that the local magnetic field in the shock region has been substantially amplified, probably by accelerated particles themselves, due to streaming instability (for recent reviews see \cite{Bykov:2013p3165,Bykov:2011p1909,Schure:2012p3068}). Despite all this circumstantial evidence, no proof has been found yet that SNRs can accelerate CRs up to the knee energy.

Charged particles can be energized at a supernova shock through diffusive shock acceleration (DSA) \cite{1977DoSSR2341306K,1978ApJ221L29B,1977ICRC11.132A,Bell:1978p1344,Bell:1978p1342}. If SNRs are the main contributors to Galactic CRs, an efficiency of $\sim 10\%$ in particle acceleration is required (see \S \ref{sec:paradigm}). The dynamical reaction of accelerated particles at a SNR shock is large enough to change the shock structure, so as to call for a non-linear theory of DSA \cite{Malkov:2001p765}. Such a theory should also be able to describe the generation of magnetic field in the shock region as due to CR-driven instabilities \cite{Amato:2006p139,Caprioli:2008p123,Caprioli:2009p157}, although many problems still need to be solved.

The combination of DSA and diffusive propagation in the Galaxy represents what I will refer to as the {\it supernova remnant paradigm}. Much work is being done at the time of this review to find solid proofs in favor or against this paradigm. I will summarize this work here. 

In this review I will summarize some recent observational results and their possible interpretations. The paper is structured as follows: in \S \ref{sec:paradigm} I summarize the main developments in the SNR paradigm for the origin of Galactic CRs. In \S \ref{sec:breaks} I briefly discuss the evidence for spectral breaks and their possible interpretations in terms of acceleration and propagation. In \S \ref{sec:balmer} I will discuss some recent developments in the study of CR acceleration in SNRs in terms of observations of Balmer lines from the shock region. A summary is provided in \S \ref{sec:conclude}.

\section{Status of the SNR paradigm}
\label{sec:paradigm}

The SNR paradigm is based on the fact that CRs may be accelerated in SNRs with an efficiency of the order of $\sim 10\%$ through diffusive shock acceleration (DSA). CRs accelerated in SNRs reach the Earth after energy dependent diffusive propagation in the Galaxy, which results in a ratio of secondary to primary nuclei (for instance B/C) that decreases with energy per nucleon at $E/n>10$ GeV/n. In this energy region the scaling with energy can be easily shown to be the same as $X(E)$ where $X(E)\propto 1/D(E)$ is the grammage and $D(E)$ is the diffusion coefficient for particles with energy $E$. If one writes $D(E)\propto E^{\delta}$, the slope of the B/C ratio tells us the value of $\delta$, and the normalization of the B/C ratio tells us about the grammage at the reference energy. In general, for relativistic energies of the propagating particles, diffusive transport of particles injected with a spectrum $Q(E)\propto E^{-\gamma}$ leads to a spectrum at the Earth which is $n(E) \propto E^{-\gamma-\delta}$. This rule of thumb is very useful to derive several interesting constraints. 

In the context of the test particle theory of particle acceleration at a strong shock, the spectrum of accelerated particles is $Q(E)\sim E^{-2}$ (in terms of the distribution function in momentum the injection would be $Q(p) \sim p^{-4}$); since the observed spectrum has a slope $\sim 2.7$, the rule $\gamma+\delta=2.7$ implies that the required diffusion coefficient should scale as $D(E)\propto E^{0.7}$. The process of diffusive particle acceleration at a shock when the dynamical reaction of the accelerated particles is taken into account is described by the non-linear theory of DSA (NLDSA) (see \cite{Malkov:2001p765} for a review). The main predictions of NLDSA are the following: 1) the spectrum of accelerated particles is no longer a power law, and becomes concave, namely steeper than $p^{-4}$ at low energies ($p\lesssim 10 m_{p}c$) and harder than $p^{-4}$ at high energies (namely harder than $E^{-2}$ in terms of kinetic energy). 2) Conservation of energy and momentum across the shock, including accelerated particles, implies that since part of the ram pressure of the upstream plasma gets converted into accelerated particles, there is less energy available for conversion into thermal energy of the background plasma, therefore the temperature of the downstream gas is lower than expected in the absence of accelerated particles. 3) The super-alfvenic drift of accelerated particles with the shock results in several CR-induced plasma instabilities, that may facilitate the process of particle acceleration by creating the scattering centers necessary to shorten the acceleration time and reach higher energies. In the absence of this phenomenon it is easy to show that the maximum energy achievable in a SNR is in the GeV range, rather than close to the knee, as observations require. 

\begin{figure}
\includegraphics[width=0.4\textwidth]{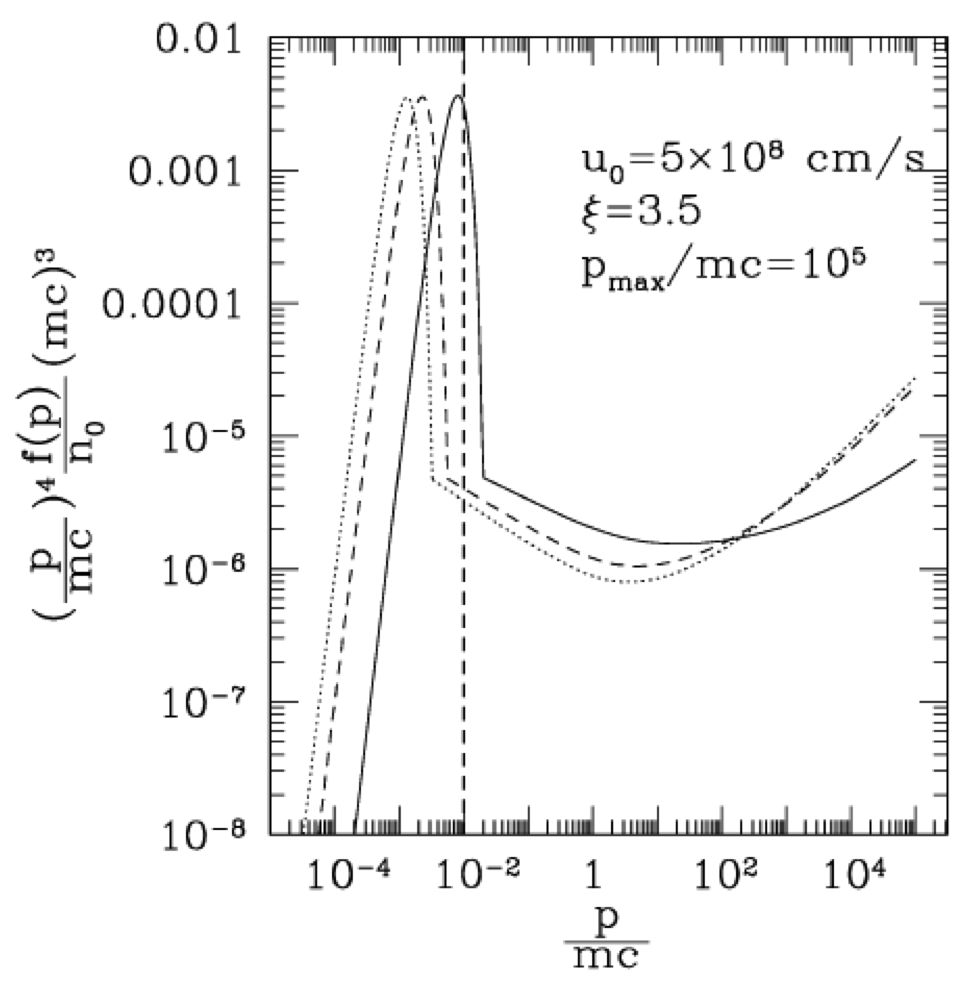}
\caption{Particle spectra (thermal plus non-thermal) at a CR modified shock with Mach number $M_{0}=10$ (solid line), $M_{0}=50$ (dashed line) and $M_{0}=100$ (dotted line). The vertical dashed line is the location of the thermal peak as expected for an ordinary shock with no particle acceleration (this value depends very weakly on the Mach number, for strong shocks). The plasma velocity at upstream infinity is $u_{0}=5\times 10^{8}$ $cm/s$, $p_{max}=10^{5} m_{p}c$ and the injection parameter is $\xi=3.5$ \cite{Blasi:2005p107}.}
\label{fig:modshockspec}      
\end{figure}

The first two effects are illustrated well in Fig. \ref{fig:modshockspec} (from \cite{Blasi:2005p107}) where I show the spectrum (thermal plus non-thermal) of the particles in the shock region for a shock Mach number $M_{0} = 10$ (solid line), $M_{0} = 50$ (dashed line) and $M_{0} = 100$ (dotted line). The thermal component is assumed to have a Maxwellian shape. Increasing the Mach number of the shock the acceleration efficiency is shown to increases, so that the spectra become increasingly more concave (the dynamical reaction of accelerated particles increases). At the same time, the peak of the thermal distribution moves leftward, namely the background plasma becomes colder for larger CR acceleration efficiency. The vertical dashed line in Fig. \ref{fig:modshockspec} shows the position of the thermal peak for a shock that does not accelerate CRs. 

One can appreciate that the spectrum of accelerated particles at high momenta becomes harder than $p^{-4}$ (namely harder than $E^{-2}$). It follows that the value of $\delta$ required to fit observations when NLDSA is used is even larger than 0.7. Realistic calculations \cite{Berezhko:2007p1010} find $\delta\simeq 0.75$. 

In both cases of DSA and NLDSA the injection spectrum is expected to be at least as hard as $E^{-2}$ at high energy. The correspondingly large value of $\delta$ implies an exceedingly large anisotropy \cite{Ptuskin:2006p666,Blasi:2012p2024} which suggests that probably our descriptions of either particle acceleration or anisotropy are unsatisfactory. 

In the last few years several SNRs have at last been detected in gamma rays, often from the GeV to the TeV energy range. This gamma ray emission is the result of inverse Compton scattering of electrons and pion production in inelastic hadronic collisions. The latter process has been viewed for a long time as the smoking gun of CR acceleration in SNRs. \cite{Caprioli:2011p2134} pointed out that the CR spectra inferred from gamma ray observations of a sample of SNRs is substantially steeper than the prediction of NLDSA (and of DSA as well). This finding suggests that the problem discussed above might be in our understanding of the process of particle acceleration. Indeed, we are aware of several effects that should go in the direction of making spectra of CRs injected by SNRs into the ISM steeper. We discuss these effects below, but it is important to keep in mind that none of these effects is currently totally under control from the theoretical point of view. 

If the velocity of the scattering centers \cite{Ptuskin:2010p1025,Caprioli:2010p133} responsible for particles' diffusion around the shock surface is large enough, the spectra of accelerated particles can become appreciably steeper, since the relevant compression factor is not the ratio of the upstream and downstream fluid velocities, but rather the ratio of the upstream and downstream velocity of the scattering centers, as seen in the shock frame. This effect depends on the helicity of the waves responsible for the scattering, and may in principle lead to a hardening of the spectrum rather than a steepening. In turn the helicity of the waves depends on numerous aspects of wave production which are very poorly known and are hardly accessible observationally. 

\begin{figure}
\includegraphics[width=0.4\textwidth]{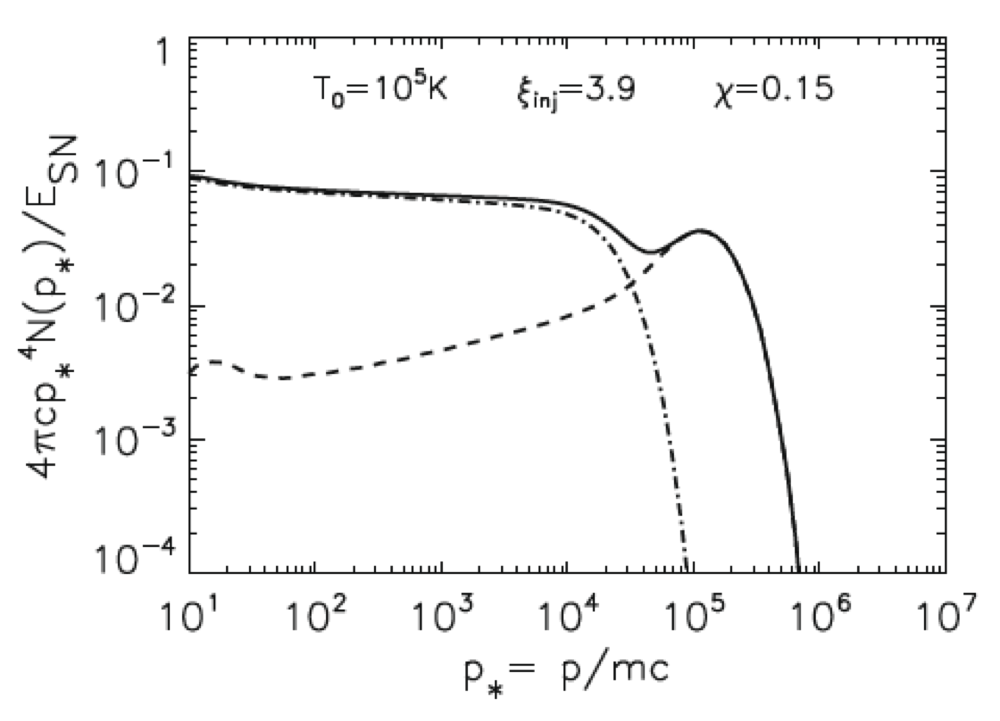}
\caption{ CR spectrum injected in the ISM by a SNR expanding in a medium with density $n_{0}=0.1~cm^{-3}$, temperature $T_{0}=10^{5}$ K and injection parameter $\xi_{inj}=3.9$ (from \cite[]{Caprioli:2010p133}). The dashed line shows the escape of particles from upstream, the dash-dotted line is the spectrum of particles escaping at the end of the evolution. The solid line is the sum of the two. The escape boundary is located at $0.15 R_{sh}$.}
\label{fig:escapeflux}      
\end{figure}

A very important point to keep in mind is that the spectrum of accelerated particles as calculated at any given time at a SNR shock is not the same as the spectrum of particles leaving the remnant to become CRs. The process that connects the accelerated particles with CRs is the escape of particles from the accelerator \cite{2011MNRAS.415.1807D}. The understanding of this process is one of the most challenging aspects of the SNR paradigm. During the Sedov-Taylor phase of the expansion of a SNR, particle escape can only occur at momenta close to the maximum momentum reached at that given time. For an observer outside the remnant the escape flux is strongly peaked at $p_{max}(t)$. On the other hand, this escape flux becomes close to a power law $\sim E^{-2}$ once it is integrated in time \cite{Caprioli:2010p133,Ptuskin:2010p1025}, and its slope has no connection with the slope expected from DSA. Non-linear effects make the spectrum of escaping particles somewhat harder than $E^{-2}$. The accelerated particles that are advected downstream of the shock lose energy adiabatically and can leave the remnant only when the SNR shock disappears. The total escape flux from a SNR was calculated in the context of NLDSA \cite{Caprioli:2010p133,Ptuskin:2010p1025} and leads to spectra such as the one in Fig. \ref{fig:escapeflux}, where the dashed line is the time-integrated escape spectrum from upstream and the dash-dotted line shows the flux of particles escaped after the shock disappearance, again integrated over the history of the remnant. 

One can notice that the overall spectrum is somewhat steeper than $E^{-2}$ although the spectrum calculated at any given time would be concave. It is also important to appreciate the importance of the escape from upstream: in the absence of this phenomenon the escape flux would have a cutoff at much lower energies (dot-dashed line), because of adiabatic losses during the expansion of the remnant. The fact that the escape flux is steeper than the instantaneous spectra alleviates but does not solve the problem of the anisotropy discussed above. Moreover, since the CR spectrum inferred from gamma ray observations is very close to the instantaneous spectrum of accelerated particles, the above considerations on escape do not affect the problem of the comparison of NLDSA with gamma ray observations \cite{Caprioli:2011p2134}.

As mentioned above, arguably the most crucial aspect of NLDSA is the phenomenon of magnetic field amplification. There are two aspects of this problem: 1) magnetic field amplification is required in order to explain the thin non-thermal X-ray rims observed in virtually all young SNRs; 2) the phenomenon is also likely to be related with the need for enhanced scattering of particles close to the shock surface. If to use only the pre-existing ISM turbulence, the maximum energy of accelerated particles would be bound to be exceedingly low, and certainly much lower than the knee energy. 

There are ways to amplify the magnetic field (possibly providing an explanation of the X-ray rims) while not appreciably affecting the confinement time of particles at the shock. For instance, magnetic field amplification can be due to plasma related phenomena \cite{Giacalone:2007p962} if the shock propagates in an inhomogeneous medium with density fluctuations $\delta\rho/\rho\sim 1$. While crossing the shock surface these inhomogeneities lead to shock corrugation and to the development of eddies in which magnetic field is frozen. The twisting of the eddies may lead to magnetic field amplification on time scales $\sim L_{c}/u_{2}$, where $L_{c}$ is the spatial size of these regions with larger density and $u_{2}$ is the plasma speed downstream of the shock. Smaller scales also grow so as to form a power spectrum downstream. This phenomenon could well be able to account for the observed thin X-ray filaments. The acceleration time for particles at the shock is however not necessarily appreciably reduced in that no field amplification occurs upstream of the shock. It turns out that this mechanism may be effective in accelerating particles in the cases where the initial magnetic field is perpendicular to the shock normal. It seems unlikely however that this scenario, so strongly dependent upon the geometry of the system, may lead to a general solution of how to reach the highest energies in Galactic CRs, although this possibility definitely deserves more attention. 

A more interesting possibility, that may address at the same time the problem of magnetic field amplification and that of making particle acceleration faster, is related to CR induced magnetic field amplification. 

It has been known for quite some time that the super-Alfv\'enic streaming of charged particles in a plasma leads to the excitation of an instability \cite{Skilling:1975p2165}. The role of this instability in the process of particle acceleration in SNR shocks was recognized and its implications were discussed by many authors, most notably \cite{Zweibel:1979p2308} and \cite{Achterberg:1983p2080}. The initial investigation of this instability led to identify as crucial the growth of resonant waves with wavenumber $k=1/r_{L}$, where $r_{L}$ is the Larmor radius of the particles generating the instability. The waves are therefore generated through the collective effect of the streaming of CRs but can be resonantly absorbed by individual particles thereby leading to their pitch angle diffusion. The resonance condition, taken at face value, would lead to expect that the growth stops when the turbulent magnetic field becomes of the same order as the pre-existing ordered magnetic field $\delta B\sim B_{0}$, so that the saturation level of this instability has often been assumed to occur when $\delta B/B\sim 1$. \cite{Lagage:1983p1347,Lagage:1983p1348} used this fact to conclude that the maximum energy that can possibly be reached in SNRs when the accelerated particles generate their own scattering centers is $\lesssim 10^{4}-10^{5}$ GeV/n, well below the energy of the knee. Hence, though the streaming instability leads to an appealing self-generation of the waves responsible for particle diffusion, the intrinsic resonant nature of the instability would inhibit the possibility to reach sufficiently high energy. It is important to notice that the problem with this instability is not the time scale, but again the resonant nature that forces $\delta B/B\sim 1$.

For simplicity let us consider the case of a spectrum of accelerated particles coincident with the canonical DSA spectrum $f_{CR,0}(p)\propto p^{-4}$ for $\gamma_{min} \leq p/m_{p}c \leq \gamma_{max}$. When the CR efficiency is small, namely when the condition
\be
\frac{n_{CR}}{n_{i}} \ll \frac{v_{A}^{2}}{V_{sh}c}
\label{eq:condition}
\ee
is fulfilled \cite{Zweibel:1979p2308,Achterberg:1983p2080}, it is easy to show that Alfv\'en waves are excited (namely $Re\left[\omega\right]\approx k v_{A}$) and their growth rate is:
\be
Im\left[\omega\right](k)  \equiv \omega_{I} (k)= \frac{\pi}{8} \Omega_{p}^{*} \frac{V_{sh}}{v_{A}}\frac{n_{CR}(p>p_{res}(k))}{n_{i}}.
\label{eq:Im}
\ee
As an order of magnitude the density of CRs can be related to the efficiency of CR acceleration as $\frac{n_{CR}}{n_{i}} \approx \frac{3\xi_{CR}}{\gamma_{min}\Lambda}\left( \frac{V_{sh}}{c}\right)^{2}$,
where $\gamma_{min}\sim 1$ is the minimum Lorentz factor of accelerated particles and $\Lambda=\ln(\gamma_{max}/\gamma_{min})$, $\gamma_{max}$ being the maximum Lorentz factor. Eq. \ref{eq:condition} becomes then
\be
\xi_{CR} \ll \frac{\gamma_{min}\Lambda}{3} \left( \frac{v_{A}}{V_{sh}}\right)^{2}\frac{c}{V_{sh}}\approx 8\times 10^{-4} \left( \frac{V_{sh}}{5\times 10^{8} cm/s}\right)^{-3},
\label{eq:condition1}
\ee
which is typically much smaller than the value $\xi_{CR}\sim 10\%$ which is required of SNRs to be the sources of the bulk of Galactic CRs. It follows that in phases in which the SNR accelerates CRs most effectively the growth rate proceeds in a different regime. 

In such regime, that occurs when Eq. \ref{eq:condition} is not fulfilled, the solution of the dispersion relation for $k r_{L,0}\leq 1$, namely for waves that can resonate with protons in the spectrum of accelerated particles ($\gamma\geq \gamma_{min}$) becomes:
\be
\omega_{I} \approx \omega_{R} =  \left[ \frac{\pi}{8} \Omega_{p}^{*} k V_{sh} \frac{n_{CR}(p>p_{res}(k))}{n_{i}} \right]^{1/2}.
\ee
Since $n_{CR}(p>p_{res}(k))\propto p_{res}^{-1}\sim k$, it follows that $\omega\propto k$ for $k r_{L,0}\leq 1$, but the phase velocity of the waves $v_{\phi}=\omega_{R}/k\gg v_{A}$. The fact that the phase velocity of these waves exceeds the Alfv\'en speed may affect the slope of the spectrum of particles accelerated at the shock, as discussed above. 

Even neglecting damping, and requiring that the saturation of the process of magnetic field amplification is solely due to the finite advection time, one can easily show that in both regimes described above at most one can achieve $\delta B/B_{0}\sim 1$, that falls short of solving the problem of reaching the knee by more than one order of magnitude. 

\cite{Bell:2004p737,2005MNRAS.358.181B} noticed that when the condition in Eq. \ref{eq:condition} is violated, namely when 
\be
\xi_{CR} > \frac{\gamma_{min}\Lambda}{3} \left( \frac{v_{A}}{V_{sh}}\right)^{2}\frac{c}{V_{sh}},
\ee
the right hand polarized mode develops a non-resonant branch for $kr_{L,0}>1$ (spatial scales smaller than the Larmor radius of all the particles in the spectrum of accelerated particles), with a growth rate that keeps increasing proportional to $k^{1/2}$ and reaches a maximum for 
\be
k_{*}r_{L,0} = \frac{3\xi_{CR}\gamma_{min}}{\Lambda}\left( \frac{V_{sh}}{v_{A}}\right)^{2}\frac{V_{sh}}{c}>1,
\ee
which is a factor $(k_{*}r_{L,0})^{1/2}$ larger than the growth rate of the resonant mode at $k r_{L,0}=1$. This non-resonant mode has several interesting aspects: first, it is current driven, but the current that is responsible for the appearance of this mode is the return current induced in the background plasma by the CR current. The fact that the return current is made of electrons moving with respect to protons is the physical reason for these modes developing on small scales (electrons in the background plasma have very low energy) and right-hand polarized. Second, the growth of these modes, when they exist, is very fast for high speed shocks, however they cannot resonate with CR particles because their scale is much smaller than the Larmor radius of any particles at the shock. On the other hand, it was shown that the growth of these modes leads to the formation of complex structures: flux tubes form, that appear to be organized on large spatial scales \cite{2012MNRAS.419.2433R} and ions are expelled from these tubes thereby inducing the formation of density perturbations. 

The problem of particle acceleration at SNR shocks in the presence of small scale turbulence generated by the growth of the non-resonant mode was studied numerically in \cite{2008ApJ.678.255Z}, where maximum energies of the order to $10^{5}$ GeV were found, as a result of the fact that at the highest energies the scattering proceeds in the small deflection angle regime $D(p)\propto p^{2}$. This finding reflects the difficulty of small scale waves to resonate with particles, irrespective of how fast the modes grow. 

Recently it was proposed that the growth of the fast non-resonant mode may in fact also enhance the growth of waves with $k r_{L,0}<1$ \cite{Bykov:2009p3106,2011MNRAS.410.39B}. If this process were confirmed by numerical calculations of the instability (current calculations are all carried out in the quasi-linear regime), it might provide a way to overcome the problem of inefficient scattering of accelerated particles off the existing turbulence around SNR shocks. 

The non-linear development of CR induced magnetic field amplification is likely to be much more complex than illustrated so far. While there is no doubt that the small scale non-resonant instability \cite{Bell:2004p737} is very fast, provided the acceleration efficiency is large enough, the question of what happens to these modes while they grow remains open. Both MHD simulations \cite{Bell:2004p737} and Particle-in-Cell simulations of this instability \cite{2009ApJ.694.626R} show how the growth leads to the development of modes on larger spatial scales. In recent numerical work \cite{2012MNRAS.419.2433R,2013ApJ.765L.20C} it has been shown that the current of CRs escaping the system induces the formation of filaments: the background plasma inside such filaments gets expelled from the filaments because of the $\vec J \times \vec B$ force. Different filaments attract each other as two currents would and give rise to filaments with larger cross section. Interestingly this instability, that might be a natural development of the Bell's instability to a strongly non-linear regime, leads to magnetic field amplification on a spatial scale comparable with the Larmor radius of particles in the CR current. However, since the current is made of particles that are trying to escape the system, the instability leads to a sort of self-confinement. The picture that seems to be arising consists in a possibly self-consistent scenario in which the highest energy particles (whichever that may be) generate turbulence on the scale of their own Larmor radius, thereby allowing particles of the same energy to return to the shock and sustain DSA \cite{2013MNRAS.431.415B,2013MNRAS.430.2873R}.

In Ref. \cite{2013MNRAS.431.415B} the authors estimated the current of particles escaping at $p_{max}$ as a function of the shock velocity and concluded that the rate of growth of the instability is such as to allow young SNRs to reach $\sim 200$ TeV energies for shock velocity $V_{sh}\sim 5000$ km/s (typical of SNRs such as Tycho), falling short of the knee by about one order of magnitude. A possible conclusion of this study might be that SNRs with an even larger velocity (therefore much younger) may be responsible for acceleration of PeV CRs. The issue of whether such young SNRs may have plowed enough material (and therefore accelerated enough particles) to account for the actual fluxes of CRs observed at Earth remains to be addressed. It is worth recalling that the argument discussed above, if applied to scenarios involving SNe type Ib,c where it has been speculated that the maximum rigidity may be as high as $\sim 10^{17}$ V \cite{Ptuskin:2010p1025}, imply considerably lower maximum energies. Future detection of CR protons of Galactic origin in such high energy region would be hardly reconcilable with DSA in SNRs of any type. 

Recent gamma ray observations of SNRs have provided us with a powerful test of the SNR paradigm for the origin of CRs. There is no lack of evidence of CR proton acceleration in SNRs close to molecular clouds (MC), that act as a target for hadronic interactions resulting in pion production. Recently the AGILE \cite{2011ApJ.742L.30G,2010A&A.516L11G,2011ApJ.742L.30G} and Fermi-LAT \cite{2010ApJ.718.348A,Ackermann:2013p3110,2010ApJ.712.459A,2010Sci.327.1103A,2009ApJ.706L.1A} collaborations claimed the detection of the much sought-after pion bump in the gamma ray spectrum. This spectral feature confirms that the bulk of the gamma ray emission in these objects is due to $pp\to \pi^{0} \to 2\gamma$. These cases, besides confirming the existing accelerated hadrons, are very important indicators of CR propagation around the sources \cite{2007Ap&SS.309.365G,2009MNRAS.396.1629G,2008ApJ...689..213R,Nava:2013p3140,2013PhRvD.88b3010G}. On the other hand, not much can be learnt from SNRs close to MCs on the acceleration of the bulk of CRs in the Galaxy, since such SNRs are usually not young and the maximum energies not very high.  

In this perspective, the cases of young individual SNRs are more instructive. The first clear detection of TeV gamma ray emission from a relatively young SNR came from the SNR RXJ1713.7-3946 \cite{2004Natur.432.75A,2006A&A.449.223A,2007A&A.464.235A}, later followed by the detection of the same remnant in the GeV energy range with the Fermi-LAT telescope \cite{2011ApJ.734.28A}. Here I will briefly discuss this case because it is instructive of how the comparison of theoretical predictions with data can drive our understanding of the acceleration environment. 

A discussion of the implications of the TeV data, together with the X-ray data on spectrum and morphology was presented in \cite{Morlino:2009p140}. A hadronic origin of the gamma ray emission would easily account for the bright X-ray rims (requiring a magnetic field of $\sim 160\mu G$), as well as for the gamma ray spectrum.  If electrons were to share the same temperature as protons, the model would predict a powerful thermal X-ray emission, which is not detected. Rather than disproving this possibility, this finding might be the confirmation of the expectation that at fast collisionless shocks electrons fail to reach thermal equilibrium with protons. In fact, the Coulomb collision time scale for this remnant turns out to exceed its age. On the other hand, it was pointed out in \cite{2010ApJ.712.287E} that even a slow rate of Coulomb scattering would be able to heat electrons to a temperature $\gtrsim 1$ keV, so that oxygen lines would be excited and they would dominate the thermal emission. These lines are not observed, thereby leading to a severe upper limit on the density of gas in the shock region, that would result in a too small pion production. \cite{2010ApJ.712.287E} concluded that the emission is of leptonic origin. This interpretation appears to be confirmed by Fermi-LAT data, that show a very hard gamma ray spectrum, incompatible with an origin related to pion production and decay. Clearly this does not necessarily mean that CRs are not efficiently accelerated in this remnant. It simply implies that the gas density in this remnant is too low for efficient pp scattering to contribute to the gamma ray spectrum. 

However, it should be pointed out that models based on ICS of high energy electrons are not problem free: first, as pointed out in \cite{Morlino:2009p140}, the density of IR light necessary to explain the HESS data as the result of ICS is $\sim 25$ times larger than expected. Second, the ICS interpretation requires a weak magnetic field of order $\sim 10 \mu G$, incompatible with the observed X-ray rims. Finally, recent data on the distribution of atomic and molecular hydrogen around SNR RXJ1713.7-3946 \cite{2012ApJ.746.82F} suggest a rather good spatial correlation between the distribution of this gas and the TeV gamma ray emission, which would be easier to explain if gamma rays were the result of $pp$ scattering. In conclusion, despite the fact that the shape of the spectrum of gamma rays would suggest a leptonic origin, the case of SNR RXJ1713.7-3946 will probably turn out to be one of those cases in which the complexity of the environment around the remnant plays a crucial role in determining the observed spectrum. Future high resolution gamma ray observations, possibly with the Cherenkov telescope array (CTA), will contribute to clarify this situation. 

A somewhat clearer case is that of the Tycho SNR, the leftover of a SN type Ia exploded in a roughly homogeneous ISM, as confirmed by the regular circular shape of the remnant. Tycho is one of the historical SNRs, as it was observed by Tycho Brahe in 1572. The multifrequency spectrum of Tycho extends from the radio band to gamma rays, and a thin X-ray rim is observed all around the remnant. It has been argued that the spectrum of gamma rays observed by Fermi-LAT \cite{2012ApJ744L2G} in the GeV range and by VERITAS \cite{2011ApJ730L20A} in the TeV range can only be compatible with a hadronic origin \cite{Morlino:2012p2243}. The morphology of the X-ray emission, resulting from synchrotron radiation of electrons in the magnetic field at the shock, is consistent with a magnetic field of $\sim 300 \mu G$, which implies a maximum energy of accelerated protons of $\sim 500$ TeV. A hadronic origin of the gamma ray emission has also been claimed by \cite{2013ApJ76314B}, where however the steep gamma ray spectrum measured from Tycho is attributed to an environmental effect: the low energy flux would be boosted because the shock is assumed to encounter small dense clumps of material. The flux of high energy gamma rays is not affected because the clumps are assumed to have a small size compared with the scattering length of the protons responsible for the production of TeV gamma rays. In the calculations of \cite{Morlino:2012p2243} the steep spectrum is instead explained as a result of NLDSA in the presence of waves moving with the Alfv\'en velocity calculated in the amplified magnetic field. In this latter case the shape of the spectrum is related, though in a model dependent way, to the strength of the amplified magnetic field, which is the same quantity relevant to determine the X-ray morphology. In the former model the steep spectrum might not be found in another SNR in the same conditions, in the absence of the small scale density perturbations assumed by the authors. 

\begin{figure*}
\includegraphics[width=220pt]{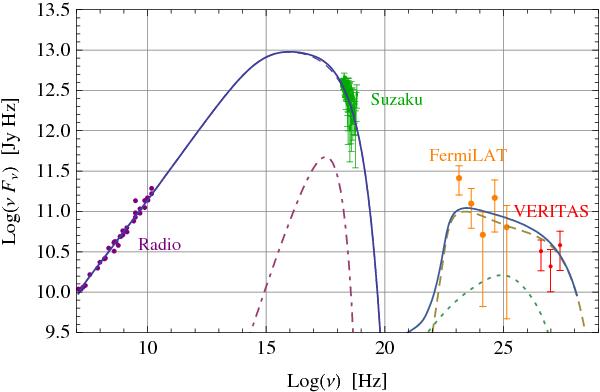}
\includegraphics[width=215pt]{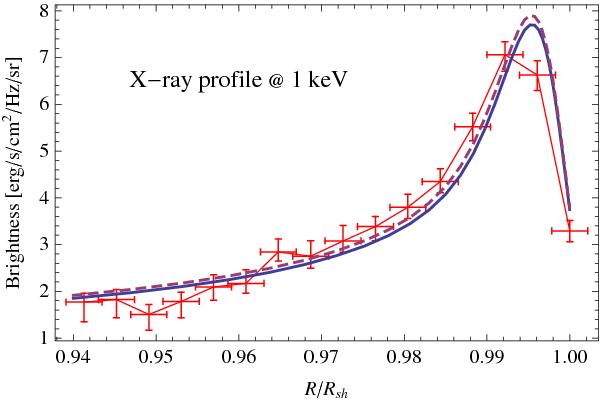}
\caption{{\it Left Panel:} Spatially integrated spectral energy distribution of Tycho. The curves show synchrotron emission, thermal electron bremsstrahlung and pion decay as calculated by \cite{Morlino:2012p2243}. Gamma ray data from Fermi-LAT \cite{2012ApJ744L2G} and VERITAS \cite{2011ApJ730L20A} are shown. {\it Right Panel:} Projected X-ray brightness at 1 keV. Data points are from \cite{2007ApJ...665..315C}. The solid line shows the result of the calculations by \cite{Morlino:2012p2243} after convolution with the Chandra point spread function.}
\label{fig:tycho}      
\end{figure*}

The multifrequency spectrum of Tycho (left) and the X-ray brightness of its rims (right) are shown in Fig. \ref{fig:tycho} (from \cite{Morlino:2012p2243}). The dash-dotted line in the left panel shows the thermal emission from the downstream gas (here the electron temperature is assumed to be related to the proton temperature as $T_{e}=(m_{e}/m_{p}) T_{p}$ immediately behind the shock, and increases with time solely due to Coulomb scattering, that couples electrons with the warmer protons), the short-dashed line shows the ICS contribution to the gamma ray flux, while the dashed line refers to gamma rays from pion decays. The solid lines show the total flux. The figure shows rather impressively how the magnetic field necessary to describe the radio and X-ray radiation as synchrotron emission also describes the thickness of the X-ray rims (right panel) and pushes the maximum energy of accelerated particles to $\sim 500$ TeV (in the assumption of Bohm diffusion). 

The case of Tycho is instructive as an illustration of the level of credibility of calculations based on the theory of NLDSA: the different techniques agree fairly well (see \cite{2010MNRAS.407.1773C} for a discussion of this point) as long as only the dynamical reaction of accelerated particles on the shock is included. When magnetic effects are taken into account, the situation becomes more complex: in the calculations based on the semi-analytical description of \cite{Amato:2006p139} the field is estimated from the growth rate and the dynamical reaction of the magnetic field on the shock is taken into account \cite{Caprioli:2008p123,Caprioli:2009p157}. Similar assumptions are adopted in \cite{2008ApJ.688.1084V}, although the technique is profoundly different. Similar considerations hold for \cite{Ptuskin:2010p1025}. On the other hand, \cite{2013ApJ76314B} take the magnetic field as a parameter of the problem, chosen to fit the observations, and its dynamical reaction is not included in the calculations. The magnetic backreaction, as discussed by \cite{Caprioli:2008p123,Caprioli:2009p157} comes into play when the magnetic pressure exceeds the thermal pressure upstream, and leads to a reduction of the compression factor at the subshock, namely less concave spectra. Even softer spectra are obtained if one introduces a recipe for the velocity of the scattering centers \cite{Ptuskin:2010p1025,Caprioli:2010p133,Morlino:2012p2243}. This, yet speculative, effect is not included in any of the other approaches.

Even more pronounced differences arise when environmental effects are included. The case of Tycho is again useful in this respect: the predictions of the standard NLDSA theory would not be able to explain the observed gamma ray spectrum from this SNR. But assuming the existence of {\it ad hoc} density fluctuations, may change the volume integrated gamma ray spectrum as to make it similar to the observed one \cite{2013ApJ76314B}. Space resolved gamma ray observations would help clarify the role of these environmental effects in forging the gamma ray spectrum of a SNR. 

\section{Spectral breaks}
\label{sec:breaks}

Changes in the shape of the spectra of CRs as observed at Earth are a promising way to gather interesting information on the physics of acceleration and propagation. This justified the great interest in the recent data from the PAMELA and CREAM experiments \cite{Adriani:2011p1893,Ahn:2010p624} that provide evidence for a change of slope in the spectra of protons and helium nuclei at rigidity $\sim 200$ GV. More specifically, PAMELA finds that the spectrum of protons between 80 and 232 GV has a slope $\gamma_{80-232} = 2.85\pm 0.015 (stat)\pm 0.004 (syst)$, while in the range $>232$ GV the slope becomes $\gamma_{>232} = 2.67\pm 0.03 (stat)\pm 0.05 (syst)$. For helium nuclei, $\gamma_{80-240} = 2.766\pm 0.01 (stat)\pm 0.027 (syst)$ and $\gamma_{>240} = 2.477\pm 0.06 (stat)\pm 0.03 (syst)$. The high energy slopes appear to be in agreement with those measured by CREAM in the TeV energy range. Two conclusions can be drawn: first, both spectra appear to have a break at rigidity $\sim 230-240$ GV. Second, the spectrum of helium is systematically harder than the proton spectrum. During this conference, the AMS-02 collaboration presented preliminary data that do not confirm the evidence for the spectral break, while confirming that the spectrum of helium is harder than that of protons. No quantitative assessment of these statements can be made at the present time since the AMS-02 data have not been published at the time of writing this review paper. 

Independent support to the slope of the proton spectrum measured by PAMELA came from the analysis of the gamma ray spectrum of gamma rays from molecular clouds in the Gould belt \cite{Neronov:2012p2217,Kachelrie:2012p10950}. Since the density in these clouds is very large, the observed gamma ray emission is mainly due to inelastic pp scattering with pion production. The spectrum of the parent CRs can be obtained by correcting for the effect of the cross section and the slope of the proton spectrum that the authors obtained between 10 and 200 GeV is compatible with the one measured by PAMELA in the same energy region. In addition, \cite{Neronov:2012p2217,Kachelrie:2012p10950} find evidence for a low energy flattening of the interstellar proton spectrum at $E\lesssim 10$ GeV. These findings are also qualitatively confirmed by the analysis of the Fermi-LAT data of the Galactic diffuse gamma ray background \cite{dermer}.

In \cite{2012MNRAS4211209T,2013arXiv13041400T} it was suggested that a local source of CRs might manifest itself in the total spectrum as a spectral hardening. This possibility becomes more realistic when the diffusion coefficient has a fast variation with energy, since in that case the fluctuations in the spectrum become sizeable (e.g. see \cite{Blasi:2012p2051}). However, these are also the situations that correspond to a large anisotropy: whenever a source contributes a flux comparable with the sum of all other sources, the anisotropy becomes of order unity \cite{Blasi:2012p2024}. This is the reason why for $\delta>0.5$ the anisotropy is usually in excess of observations at sufficiently high energy. 

In \cite{ptuskin} it was suggested that the observed spectra may reflect the efficient CR acceleration at the forward and reverse shock of young SNRs. The sum of the two, with the concavity induced by the CR dynamical reaction may produce spectra that resemble the ones that have been observed by PAMELA. However, one should keep in mind that for the concave spectra that are predicted in the context of NLDSA, the observed spectra imply that the diffusion coefficient in the Galaxy should have a rather fast dependence on energy, typically $D(E)\sim E^{0.75}$, so that again one expects strong anisotropy in this model, contrary to what is observed. 

The fact that the PAMELA spectral feature appears at the same rigidity for protons and helium is suggestive that it may be due to propagation effects. Two explanations have been put forward that are based on subtle aspects of CR propagation. The first is that the diffusion coefficient can be space and energy dependent in a non-separable way, thereby causing a spectral break \cite{tomassetti} in the CR spectrum as measured at the Earth. The second explanation is based on taking into account both the scattering in the self-generated turbulence, induced by CRs while streaming in the Galaxy, and the scattering in a pre-existing turbulence cascading towards small scales from the much larger scales where the turbulence is initially injected. In Ref. \cite{Blasi:2012p2344} the authors show that within this framework the flux of protons agrees very well with PAMELA and CREAM. It was later showed \cite{2013JCAP07001A} that the B/C ratio and other observables are also well described by this theory, which returns the energy dependence and normalization of the diffusion coefficient as well. 

\section{CR acceleration in partially ionized media and Balmer line emission in SNRs}
\label{sec:balmer}

H$\alpha$ optical emission from Balmer dominated SNR shocks is a powerful indicator of the conditions around the shock \cite{1978ApJ.225L.27C,1980ApJ.235.186C} including the presence of accelerated particles (see \cite{2010PASA.27.23H} for a review). The H$\alpha$ line is produced when neutral hydrogen is present in the shock region, and it gets excited by collisions with thermal ions and electrons to the level $n=3$ and decays to the level $n=2$. In the following I describe the basic physics aspects of this phenomenon and how it can be used to gather information on the CR energy content at the shock. 

A collisionless shock propagating in a partially ionized background goes through several interesting new phenomena: first, neutral atoms cross the shock surface without suffering any direct heating, due to the collisionless nature of the shock (all interactions are of electro-magnetic nature, therefore the energy and momentum of neutral hydrogen cannot be changed). However, a neutral atom has a finite probability of undergoing either ionization or a charge exchange  reaction, whenever there is a net velocity difference between ions and atoms. Behind the shock, ions are slowed down (their bulk motion velocity drops down) and heated up, while neutral atoms remain colder and faster. The reactions of charge exchange lead to formation of a population of hot atoms (a hot ion downstream catches an electron from a fast neutral), which also have a finite probability of getting excited. The Balmer line emission from this population corresponds to a Doppler broadened line with a width that reflects the temperature of the hot ions downstream. Measurements of the width of the broad Balmer line have often been used to estimate the temperature of protons behind the shock, and in fact it is basically the only method to do so, since at collisionless shocks electrons (which are responsible for the continuum X-ray emission) have typically a lower temperature than protons. Equilibration between the two populations of particles (electrons and protons) may eventually occur either collisionally (through Coulomb scattering) or through collective processes. 
The broad Balmer line is produced by hydrogen atoms that suffer at least one charge exchange reaction downstream of the shock. The atoms that enter downstream and are excited before suffering a charge exchange also contribute to the $H\alpha$ line, but the width of the line reflects the gas temperature upstream, and is therefore narrow (for a temperature of $10^{4}$ K, the width is $21$ km/s).  In summary, the propagation of a collisionless shock through a partially ionized medium leads to H$\alpha$ emission, consisting of a broad and a narrow line (see the recent review in \cite{2013SSRv.tmp.75G}). 

When CRs are efficiently accelerated at the shock, two new phenomena occur, as discussed in \S \ref{sec:paradigm}: 1) the temperature of the gas downstream of the shock is lower than in the absence of accelerated particles. 2) A precursor is formed upstream, as a result of the pressure exerted by accelerated particles. 

Both these phenomena have an impact on the shape and brightness of the Balmer line emission. The lower temperature of the downstream gas leads to a narrower broad Balmer line, whose width bears now information on the pressure of accelerated particles, through the conservation equations at the shock. 

The CR-induced precursor slows down the upstream ionized gas with respect to hydrogen atoms, which again do not feel the precursor but through charge exchange. If ions are heated in the precursor (not only adiabatically, but also because of turbulent heating) the charge exchange reactions transfer some of the internal energy to neutral hydrogen, thereby heating it. This phenomenon results in the broadening of the narrow Balmer line. 

A narrower broad Balmer line and a broader narrow Balmer line are both signatures of CR acceleration at SNR shocks \cite{2010PASA.27.23H}. The theory of CR acceleration at collisionless SNR shocks in the presence of neutral hydrogen has only recently been formulated \cite{2012ApJ.755.121B,2012p3105,2013ApJ.768.148M} and has led to the prediction of several new interesting phenomena.

The presence of neutrals in the shock region changes the structure of the shock even in the absence of appreciable amounts of accelerated particles, due to the phenomenon of neutral return flux \cite{2012ApJ.755.121B}. A neutral atom that crosses the shock and suffers a charge exchange reaction downstream gives rise to a new neutral atom moving with high bulk velocity. There is a sizeable probability (dependent upon the shock velocity) that the resulting atom moves towards the shock and crosses it towards upstream. A new reaction of either charge exchange or ionization upstream leads the atom to deposit energy and momentum in the upstream plasma, within a distance of the order of its collision length. On the same distance scale, the upstream plasma gets heated up and slows down slightly, thereby resulting in a reduction of the plasma Mach number immediately upstream of the shock (within a few pathlengths of charge exchange and/or ionization). This implies that the shock strength drops, namely its compression factor becomes less than 4 (even for strong shocks). 

This neutral return flux \cite{2012ApJ.755.121B} plays a very important role in the shock dynamics for velocity $V_{sh}\lesssim 3000$ km/s. For faster shocks, the cross section for charge exchange drops rather rapidly and ionization is more likely to occur downstream. This reduces the neutral return flux and the shock modification it produces. 

The consequences of the neutral return flux both on the process of particle acceleration and on the shape of the Balmer line are very serious: some hydrogen atoms undergo charge exchange immediately upstream of the shock, with ions that have been heated by the neutral return flux. These atoms give rise to a Balmer line emission corresponding to the temperature of the ions immediately upstream of the shock. As demonstrated in \cite{2012p3105} this contribution consists of an intermediate Balmer line, with a typical width of $\sim 100-300$ km/s. Some tentative evidence of this intermediate line might have already been found in existing data ({\it e.g.} see \cite{2000ApJ.535.266G}).

\begin{figure}
\includegraphics[width=0.4\textwidth]{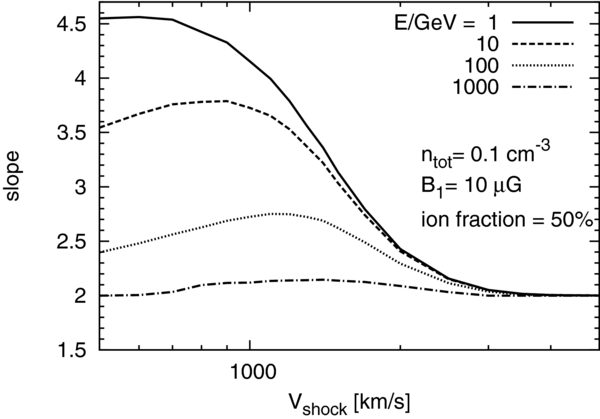}
\caption{Slope of the differential spectrum of test particles accelerated at a shock propagating in a partially ionized medium, with density $0.1~cm^{-3}$, magnetic field $10\mu G$ and ionized fraction of $50\%$, as a function of the shock velocity. The lines show the slope for particles at different energies, as indicated. The figure is taken from the paper by \cite{2012ApJ.755.121B}.}
\label{fig:slope}      
\end{figure}

The most striking consequence of the neutral return flux is however the steepening of the spectrum of test particles accelerated at the shock, first discussed in \cite{2012ApJ.755.121B}. The effect is caused by the reduction of the compression factor of the shock, which reflects on the fact that the slope of the spectrum of accelerated particles gets softer. This effect is however limited to particles that diffuse upstream of the shock out to a distance of order a few collision lengths of charge exchange/ionization upstream. It follows that the steepening of the spectrum is limited to particle energies low enough as to make their diffusion length shorter than the pathlength for charge exchange and ionization. In Fig. \ref{fig:slope} (from \cite{2012ApJ.755.121B}) I show the spectral slope as a function of shock velocity for particles with energy 1, 10, 100, 1000 GeV, as labelled (background gas density, magnetic field and ionization fraction are as indicated). One can see that the standard slope $\sim 2$ is recovered only for shock velocities $>3000$ km/s. For shocks with velocity $\sim 1000$ km/s the effect may make the spectra extremely steep, to the point that the energy content may be dominated by the injection energy, rather than, as it usually is, by the particle mass. This situation, for all practical purposes, corresponds to not having particle acceleration but rather a strong modification of the distribution of thermal particles. For milder neutral induced shock modifications, the effect is that of making the spectra of accelerated particles softer. It is possible that this effect may play a role in reconciling the predicted CR spectra with those inferred from gamma ray observations, although limited to shocks with relatively low velocity, $\lesssim 3000$ km/s. 

So far, I have limited the discussion to the case of test particle acceleration. However, it is clear that the shock modification is both due to neutrals and to the dynamical reaction of accelerated particles (\S \ref{sec:paradigm}).

The theory of NLDSA in the presence of partially ionized media was fully developed in \cite{2013ApJ.768.148M}, using the kinetic formalism introduced in \cite{2012ApJ.755.121B} to account for the fact that neutral atoms do not behave as a fluid, and their distribution in phase space can hardly be approximated as being a maxwellian. The theory describes the physics of particle acceleration, taking into account the shock modification induced by accelerated particles as well as neutrals, and magnetic field amplification. The theory is based on a mixed technique in which neutrals are treated through a Boltzmann equation while ions are treated as a fluid. The collision term in the Boltzmann equation is represented by the interaction rates of hydrogen atoms due to charge exchange with ions and ionization, at any given location. The Boltzmann equation for neutrals, the fluid equations for ions and the non-linear partial differential equation for accelerated particles are coupled together and solved by using an iterative method. The calculation returns the spectrum of accelerated particles at any location, all thermodynamical quantities of the background plasma (density, temperature, pressure) at any location, the magnetic field distribution, and the distribution function of neutral hydrogen in phase space at any location from far upstream to far downstream. 

\begin{figure*}
\includegraphics[width=.9\textwidth]{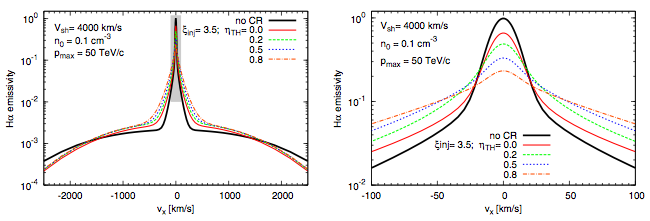}
\caption{{\it Left Panel:} Shape of the Balmer line emission for a shock moving with velocity $V_{sh}=4000$ km/s in a medium with density $0.1~cm^{-3}$, as calculated by \cite{2013ApJ.768.148M}. The thick (black) solid line shows the result in the absence of particle acceleration. The other lines show the broadening of the narrow component and the narrowing of the broad component when CR are accelerated with an injection parameter $\xi_{inj}=3.5$ and different levels of turbulent heating ($\eta_{TH}$) as indicated. {\it Right Panel:} Zoom in of the left panel on the region of the narrow Balmer line, in order to emphasize the broadening of the narrow component in the case of efficient particle acceleration.}
\label{fig:balmerline}      
\end{figure*}

These quantities can then be used to infer the Balmer line emission from the shock region, taking into account the excitation probabilities to the different atomic levels in hydrogen. An instance of such calculation is shown in Fig. \ref{fig:balmerline}, where I show the shape of the Balmer line for a shock moving with velocity $V_{sh}=4000$ km/s in a medium with density $0.1~cm^{-3}$ with a maximum momentum of accelerated particles $p_{max}=50$ TeV/c. The left panel shows the whole structure of the line, including the narrow and broad components, while the right panel shows a zoom-in on the narrow Balmer line region (gray shadowed region in the left panel). The black line is the Balmer line emission in the absence of accelerated particles. Allowing for particle acceleration to occur leads to a narrower broad Balmer line (left panel) and to a broadening of the narrow component (right panel). The latter is rather sensitive however to the level of turbulent heating in the upstream plasma, namely the amount of energy that is damped by waves into thermal energy of the background plasma. In fact turbulent heating is also responsible for a more evident intermediate Balmer line (better visible in the left panel) with a width of few hundred km/s. It is worth recalling that observations of the Balmer line width are usually aimed at either the narrow or the broad component, but usually not both, because of the very different velocity resolution necessary for measuring the two lines. Therefore the intermediate line is usually absorbed in either the broad or the narrow component, depending on which component is being measured. This implies that an assessment of the observability of the intermediate Balmer component requires a proper convolution of the predictions with the velocity resolution of the instrument. 

At the time of this review, an anomalous shape of the broad Balmer line has been reliably measured in a couple of SNRs, namely SNR 0509-67.5 \cite{Helder:2010p618,Helder:2011p2386} and SNR RCW86 \cite{Helder:2009p660}. As I discuss below, the main problem in making a case for CR acceleration is the uncertainty in the knowledge of the shock velocity and the degree of electron-ion equilibration downstream of the shock. The ratio of the electron and proton temperatures downstream is indicated here as $\beta_{down}=T_{e}/T_{p}$. The other parameters of the problem have a lesser impact on the inferred value of the CR acceleration efficiency. 

The SNR 0509-67.5 is located in the Large Magellanic Cloud (LMC), therefore its distance is very well known, $50\pm 1$ kpc. \cite{Helder:2010p618,Helder:2011p2386} carried out a measurement of the broad component of the H$\alpha$ line emission in two different regions of the blast wave of SNR 0509-67.5, located in the southwest (SW) and northeast (NE) rim, obtaining a FWHM of $2680 \pm 70$ km/s and $3900 \pm 800$ km/s, respectively. The shock velocity was estimated to be $V_{sh} = 6000 \pm 300$ km/s when averaged over the entire remnant, and $6600\pm 400$ km/s in the NE part, while a value of 5000 km/s was used by \cite{Helder:2010p618,Helder:2011p2386} for the SW rim. The width of the broad Balmer line was claimed by the authors to be suggestive of efficient CR acceleration. In order to infer the CR acceleration efficiency the authors made use of the calculations by \cite{2008ApJ.689.1089V}, that, as discussed by \cite{2013arXiv1306.6454M}, adopt some assumptions on the distribution function of neutral hydrogen that may lead to a serious overestimate of the acceleration efficiency for fast shocks. Moreover, a closer look at the morphology of this SNR, reveals that the SW rim might be moving with a lower velocity than assumed in \cite{Helder:2010p618,Helder:2011p2386}, possibly as low as $\sim 4000$ km/s. Both these facts have the effect of implying a lower CR acceleration efficiency, as found by \cite{2013arXiv1306.6762M}. 

\begin{figure*}
\includegraphics[width=1.0\textwidth]{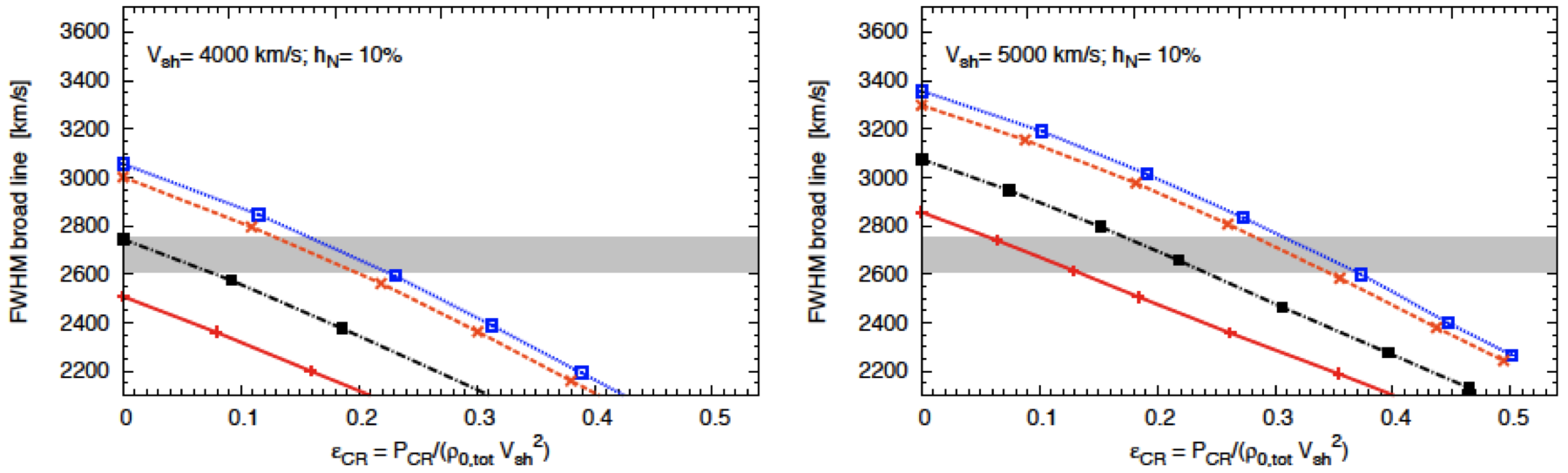}
\caption{FWHM of the broad Balmer line as a function of the CR acceleration efficiency for the SNR 0509-67.5, as calculated by \cite{2013arXiv1306.6762M}, assuming a shock velocity $V_{sh}=4000$ km/s (left panel) and $V_{sh}=5000$ km/s (right panel) and a neutral fraction $h_{N}=10\%$. The lines (from top to bottom) refer to different levels of electron-ion equilibration, $\beta_{down}=0.01,~0.1,~0.5,~1$, The shadowed region is the FWHM with $1\sigma$ error bar, as measured by \cite{Helder:2010p618}.}
\label{fig:SNR0509}      
\end{figure*}

In Fig. \ref{fig:SNR0509} (from \cite{2013arXiv1306.6762M}) I show the FWHM of the broad Balmer line in the SW rim of SNR 0509-67.5 as a function of the acceleration efficiency, for shock velocity $V_{sh}=4000$ km/s (on the left) and $V_{sh}=5000$ km/s (on the right) and a neutral fraction $h_{N}=10\%$. The shaded area represents the FWHM as measured by \cite{Helder:2010p618,Helder:2011p2386}, with a $1\sigma$ error bar. The curves refer to $\beta_{down}=0.01,~0.1,~0.5,~1$ from top to bottom. For low shock speed and for full electron-ion equilibration ($\beta_{down}=1$) the measured FWHM is still compatible with no CR acceleration. On the other hand, for such fast shocks, it is found that $\beta_{down}\ll 1$ \cite{2007ApJ.654L.69G,2013SSRv.tmp.75G}, in which case one can see that acceleration efficiencies of $\sim 10-20\%$ can be inferred from the measured FWHM. 

The case of RCW86 is more complex: the results of a measurement of the FWHM of the broad Balmer line were reported by \cite{Helder:2009p660}, where the authors claimed a FWHM of $1100\pm 63$ km/s with a shock velocity of $6000\pm 2800$ km/s and deduced a very large acceleration efficiency ($\sim 80\%$). In a more recent paper by the same authors \cite{2013arXiv1306.3994H}, the results of \cite{Helder:2009p660} were basically retracted: several regions of the SNR RCW86 were studied in detail and lower values of the shock velocity were inferred. Only marginal evidence for particle acceleration was found in selected regions. The morphology of this remnant is very complex and it is not easy to define global properties. Different parts of the SNR shock need to be studied separately. In addition, the uncertainty in the distance to SNR RCW86 is such as to make the estimate of the acceleration efficiency even more difficult. 

Anomalous widths of narrow Balmer lines have also been observed in several SNRs (see, e.g. \cite{Sollerman:2003p615}). The width of such lines is in the 30-50 km/s range, implying a pre-shock temperature around 25,000-50,000 K. If this were the ISM equilibrium temperature there would be no atomic hydrogen, implying that the pre-shock hydrogen is heated by some form of shock precursor in a region that is sufficiently thin so as to make collisional ionization equilibrium before the shock unfeasible. The CR precursor is the most plausible candidate to explain such a broadening of the narrow line.

Most important would be to have measurements of the width of the narrow and broad components (and possibly intermediate component) of the Balmer line at the same location in order to allow for a proper estimate of the CR acceleration efficiency. Co-spatial observation of the thermal X-ray emission would also provide important constraints on the electron temperature. So far, this information is not yet available with the necessary accuracy in any of the astrophysical objects of relevance. 

Recent observations of the Balmer emission from the NW rim of SN1006 \cite{Nikolic:2013p3111} have revealed a rather complex structure of the collisionless shock. That part of the remnant acts as a bright Balmer source, but does not appear to be a site of effective particle acceleration, as one can deduce from the absence of non-thermal X-ray emission from that region. This reflects in a width of the broad Balmer line that appears to be compatible with the estimated shock velocity in the same region, with no need for the presence of accelerated particles. The observations of \cite{Nikolic:2013p3111} provide however a rather impressive demonstration of the huge potential of Balmer line observations, not only to infer the CR acceleration efficiency, but also as a tool to measure the properties of collisionless shocks. 

\section{Summary}
\label{sec:conclude}

In this paper I reviewed some recent observational results and their interpretation, with focus of what we can learn in terms of CR acceleration and propagation. Our understanding of the origin of the bulk of Galactic CRs is still strongly based upon the so-called SNR paradigm, but the structure of the paradigm is constantly adapting to the wealth of data that are being collected (\S \ref{sec:paradigm}). The detection of X-ray non-thermal X-ray filaments has triggered a flood of investigations of magnetic field amplification, a phenomenon that plays a central role in the SNR paradigm as a tool to reach energies close to the knee. Despite much theoretical work is being done to understand the different mechanisms that may produce magnetic field amplification on the relevant scales, a clear understanding of how it may be possible to reach energies as high as the knee energy in SNRs is all but achieved at the present time (see discussion in \S \ref{sec:paradigm}). 

Gamma ray observations in the GeV and TeV energy band have provided us with the first direct evidence for proton acceleration in SNRs. Most evidence does in fact refer to SNRs near molecular clouds: these sources are probably more useful indicators of CR propagation around the sources than they are tools to understand acceleration. On the other hand, gamma rays from individual SNRs, such as RXJ1713.7-3946 and Tycho have taught us a great deal on acceleration of CRs, but also made us realize the limits of our theoretical formalism. In all these cases, it appears that environmental effects (for instance inhomogeneous distribution of the target gas) affects the spectrum of the observed radiation more than details of our theories. There is however a more general problem that arises from comparing the predictions of NLDSA with observations: the former leads us to predict relatively hard spectra of accelerated particles, which in turn force us to require a rather strong energy dependence of the Galactic diffusion coefficient. Such diffusion coefficient appears to be in contradiction with the relatively low observed anisotropy of high energy CRs. Moreover, statistically speaking, the spectra of gamma ray emitting SNRs seem to be too steep with respect to the basic theory of NLDSA. 

Much attention has been dedicated to studying the problem of escape of particles from SNRs, although this phenomenon remains poorly understood. A simple argument can be written down to show that the spectrum of escaping CRs (integrated in time through the history of the SNR expansion) is steeper than the instantaneous spectrum of accelerated particles. The latter is expected to reflect more directly in the spectra of the gamma ray emission from individual SNRs.

All these reasons for concern can only be properly addressed observationally by carrying out observations with as high angular resolution as possible and on a large energy span. In Gamma rays this will be possible with the upcoming Cherenkov Telescope Array, both by observing spectrum and morphology of individual SNRs as well as by studying the escape flux through the signatures it may leave on the gamma radiation from molecular clouds illuminated by CRs accelerated in a nearby SNR. Whether this will be sufficient to solve our open problems it remains to be seen. 

The detection of spectral breaks in the spectra of protons and helium nuclei by PAMELA (in agreement with what measured by CREAM at higher energies) stimulated a vast interest, as these breaks may be suggestive of new phenomena in either acceleration or propagation. In \S \ref{sec:breaks} I summarized some of the models put forward to explain these data. Unfortunately, the AMS-02 preliminary data presented during this ICRC have cast doubts on the very existence of such breaks. It will take time to clarify if the PAMELA spectral features are real or the artifact of the measurement. 

I dedicated the last part of the review (\S \ref{sec:balmer}) to discuss a novel way to look for CR acceleration in SNRs, using the widths (and intensities) of the Balmer lines produced when a SNR shock propagates in a partially ionized medium. This technique has a great potential for discovery and will hopefully inspire observers in dedicating observational time to this type of measurement.

\vspace*{0.5cm}
\footnotesize{{\bf Acknowledgment:}{The author is grateful to the Organizing Committee of the 33$^{rd}$ ICRC 2013 for support and hospitality in Rio. This work was partially funded through grant PRIN INAF 2010 and ASTRI.}

\end{document}